\documentclass[11pt]{article}
\begin{document}
\title{{\bf{\Large Quantum Tunneling and Trace Anomaly}}}
\author{
 {\bf {\normalsize Rabin Banerjee}$
$\thanks{E-mail: rabin@bose.res.in}},\, 
 {\bf {\normalsize Bibhas Ranjan Majhi}$
$\thanks{E-mail: bibhas@bose.res.in}}\\
 {\normalsize S.~N.~Bose National Centre for Basic Sciences,}
\\{\normalsize JD Block, Sector III, Salt Lake, Kolkata-700098, India}
\\[0.3cm]
}

\maketitle

{\bf Abstract:}\\
      We compute the corrections, using the tunneling formalisim based on a quantum WKB approach, to the Hawking temperature and Bekenstein-Hawking entropy for the Schwarzschild black hole. The results are related to the trace anomaly and are shown to be equivalent to findings inferred from Hawking's original calculation based on path integrals using zeta function regularization. Finally, exploiting the corrected temperature and periodicity arguments we also find the modification to the original Schwarzschild metric which captures the effect of quantum corrections.

\section{Introduction}
     ~~~~~~~~~Hawking \cite{Hawking} gave the idea that black holes are not perfectly black, rather they radiate energy continuously. Since his original calculation, several derivations of Hawking effect were subsequently presented in the literature \cite{Hartle,Gibbons,Fulling,Robinson,Iso,SK1,SK2,SK3}. A particularly intuitive method is the visualisation of the source of radiation as tunneling of s-waves \cite{Wilczek,Paddy,Berezin,Majhi,Bibhas}. Although the results in the tunneling formulation agree with Hawking's original calculation \cite{Hawking}, their connection is obscure. Also, most of the computations are only confined to the semiclassical approximation {\footnote{For a recent review and a complete list of papers, see \cite{Majhi1}.}}. It is not obvious how to go beyond this approximation and whether the results still continue to agree. Both issues are addressed here.

       In this paper we explicitly compute the corrections to the semiclassical expressions for the thermodynamic entities of the Schwarzschild black hole in the tunneling approach. An exact equivalence between our approach and Hawking's \cite{Hawking1} original calculation based on path integrals using zeta function regularization will be established.

     We briefly summarise our methodology. To begin with, we will first give a new approach to get the quantum Hamilton-Jacobi equation from the definition of the quantum canonical momentum. Then following the method suggested in \cite{Majhi1}, the corrected form of the one particle action is computed for the Schwarzschild black hole. Exploiting the ``detailed balance'' condition \cite{Paddy,Bibhas}, the modified form of the Hawking temperature is obtained and from there, using the Gibbs form of the first law of thermodynamics, the famous logarithmic and inverse powers of area corrections to the Bekenstein-Hawking area law, shown earlier in \cite{Fursaev,Majumdar,Page}, are reproduced. We find, using a constant scale transformation to the metric, that the coefficient of the logarithmic correction is related to the trace anomaly. Precisely the same result is inferred from Hawking's computations involving the one loop correction to the gravitational action \cite{Hawking1} due to fluctuations in presence of scalar fields in the spacetime. Finally, exploiting the corrected temperature and periodicity arguments \cite{Hawking2}, we derive the modifications to the original Schwarzschild black hole metric. Confining to the ${\cal{O}}(\hbar)$ contribution of this modified metric we show our results are similar to those given in the existing literature \cite{York,Lousto} obtained by including the one loop back reaction effect.

       In section 2 a quantum version of the Hamilton-Jacobi formalism is developed from which the corrected forms of the thermodynamic entities are obtained. In the leading (${\cal{O}}(\hbar)$) approximation the corrections involve logarithmic terms. Higher order corrections contain inverse powers of the black hole mass or area. All these corrections contain undetermined constants as their normalization. In section 3, we identify the normalization of the leading (logarithmic) correction with the trace anomaly. An explicit value of this normalization is thereby obtained. We also demonstrate the equivalence of our findings with Hawking's original analysis \cite{Hawking1}. Section 4 reveals, using periodicity arguments, a correction to the original Schwarzschild metric as a consequence of the modified Hawking temperature. Our concluding remarks are given in section 5.

\section{Corrected forms of Hawking temperature and entropy from quantum Hamilton-Jacobi equation}
       The tunneling method involves calculating the imaginary part of the action for the (classically forbidden) process of s-wave emission across the horizon which in turn is related to the Boltzmann factor for emission at the Hawking temperature \cite{Wilczek,Paddy,Majhi}. We consider a massless scalar particle in a general class of static, spherically symmetric spacetime of the form 
\begin{eqnarray}
ds^2=-f(r)dt^2+\frac{dr^2}{g(r)}+r^2d\Omega^2
\label{1.1}
\end{eqnarray}   
satisfying the massless Klein-Gordon equation which, in the operator form, is written as
\begin{eqnarray}
g^{\mu\nu}{\hat{p}}_\mu{\hat{p}}_\nu\phi= 0
\label{1.2}
\end{eqnarray}
where $\phi$ is some massless scalar field. Now considering the eigenvalue equation ${\hat{p}}_\mu\phi=-i\hbar\nabla_\mu\phi=p_\mu\phi$ with the classical canonical momentum $p_\mu=-\partial_\mu S$, (\ref{1.2}) is expanded as
\begin{eqnarray}
&&g^{tt}\partial_t^2S+g^{rr}\partial_r^2S-g^{tt}\Gamma^\sigma_{tt}(\partial_\sigma S)-g^{rr}\Gamma^\sigma_{rr}(\partial_\sigma S)-\frac{i}{\hbar}g^{tt}(\partial_tS)^2-\frac{i}{\hbar}g^{rr}(\partial_r S)^2=0
\label{1.5}
\end{eqnarray}
where `$\nabla_\mu$' is the covariant derivative and $S(r,t)$ is the one particle action for the scalar particle. But for the metric (\ref{1.1}) we have the following non-vanishing inverse metric coefficients and connection terms
\begin{eqnarray}
g^{tt}=-\frac{1}{f};\,\ g^{rr}=g;\,\, \Gamma^r_{tt}=\frac{f'g}{2};\,\,\  \Gamma^r_{rr}=-\frac{g'}{2g}.
\label{1.6}
\end{eqnarray}
Substituting these values in (\ref{1.5}) we obtain
\begin{eqnarray}
&&\frac{i}{\sqrt{f(r)g(r)}}\Big(\frac{\partial S}{\partial t}\Big)^2 - i\sqrt{f(r)g(r)}\Big(\frac{\partial S}{\partial r}\Big)^2 - \frac{\hbar}{\sqrt{f(r)g(r)}}\frac{\partial^2 S}{\partial t^2} + \hbar \sqrt{f(r)g(r)}\frac{\partial^2 S}{\partial r^2}
\nonumber
\\
&&+ \frac{\hbar}{2}\Big(\frac{\partial f(r)}{\partial r}\sqrt{\frac{g(r)}{f(r)}}+\frac{\partial g(r)}{\partial r}\sqrt{\frac{f(r)}{g(r)}}\Big)\frac{\partial S}{\partial r}=0
\label{1.7}
\end{eqnarray}
which is the desired quantum Hamilton-Jacobi equation. Setting $\hbar=0$ yields the usual classical Hamilton-Jacobi equation. The above relation can also be obtained by a direct substitution of the ansatz, 
\begin{eqnarray}
\phi(r,t)={\textrm{exp}}\Big[-\frac{i}{\hbar}S(r,t)\Big]
\label{ansatz}
\end{eqnarray}
in (\ref{1.2}) \cite{Majhi1}. In this sense the quantum Hamilton-Jacobi equation (\ref{1.7}) implies the quantum Klein-Gordon equation (\ref{1.2}).

      Now expanding $S(r,t)$ in powers of $\hbar$, we find,
\begin{eqnarray}
S(r,t)=S_0(r,t)+\sum_i \hbar^i S_i(r,t).
\label{1.9}
\end{eqnarray}
where $i=1,2,3,......$. In this expansion the terms from ${\cal{O}}(\hbar)$ onwards are treated as quantum corrections over the semiclassical value $S_0$. Substituting (\ref{1.9}) in (\ref{1.7}) and equating terms involving identical powers in $\hbar$, we obtain the same set of equations for all $S$'s \cite{Majhi1},
\begin{eqnarray}
\frac{\partial S_a}{\partial t}=\pm\sqrt{f(r)g(r)}\frac{\partial S_a}{\partial r};\,\, a=0,1,2,3,.....
\label{new}
\end{eqnarray}

   To obtain a solution for $S(r,t)$ we first solve for $S_0(r,t)$. Now since the metric (\ref{1.1}) is stationary, it has timelike Killing vectors and therefore we consider a solution for $S_0(r,t)$ of the form,
\begin{eqnarray}
S_0(r,t)=\omega t+\tilde{S}_0(r)
\label{tunn1}
\end{eqnarray}
where $\omega$ is interpreted as the conserved quantity associated with the timelike Killing vector. Substituting this in the first equation of the set (\ref{new}) (i.e. for $a=0$) a solution for $\tilde{S}_0$ is obtained. Inserting this back in (\ref{tunn1}) yields,
\begin{eqnarray}
S_0(r,t)=\omega t  \pm\omega\int\frac{dr}{\sqrt{f(r)g(r)}}.
\label{tunn11}
\end{eqnarray}
The $+(-)$ sign is for ingoing (outgoing) particle and the limits of the integration are chosen such that the particle goes through the event horizon $r=r_H$. Likewise, the other pieces $S_i(r,t)$ appearing in (\ref{new}) are also functions of ($t\pm r_*$) where $r_*=\int \frac{dr}{\sqrt{fg}}$. To see we take, following (\ref{tunn1}), the following ansatz,
\begin{eqnarray}
S_i(r,t)=\omega_i t+{\tilde{S}}_i(r)
\label{tunn12} 
\end{eqnarray}
Inserting this in (\ref{new}) yields,
\begin{eqnarray}
{\tilde{S}}_i(r)=\pm\omega_i r_*=\pm\omega_i\int\frac{dr}{\sqrt{fg}}
\label{tunn13}
\end{eqnarray}
Combining all the terms we obtain,
\begin{eqnarray}
S(r,t)= \Big(1+\sum_i\hbar^i\gamma_i\Big)(t  \pm r_*)\omega
\label{tunn14}
\end{eqnarray}
where $\gamma_i=\frac{\omega_i}{\omega}$. From the expression within the first parentheses of the above equation it is found that $\gamma_i$ must have dimension of $\hbar^{-i}$. Again in units of $G=c=k_B=1$ the Planck constant $\hbar$ is of the order of square of the Planck Mass $M_P$ and so from dimensional analysis $\gamma_i$ must have the dimension of $M^{-2i}$ where $M$ is the mass of the black hole. Specifically, for Schwarzschild type black holes having mass as the only macroscopic parameter, these considerations show that (\ref{tunn14}) has the form,
\begin{eqnarray}
S(r,t)=\Big(1+\sum_i\beta_i\frac{\hbar^i}{M^{2i}}\Big)(t\pm r_*)\omega
\label{1.10}
\end{eqnarray}
where $\beta_i$'s are dimensionless constant parameters. 
Now keeping in mind that the ingoing probability $P_{(\textrm {in})}=|\phi_{(\textrm {in})}|^2$ has to be unity in the classical limit (i.e. $\hbar\rightarrow 0$), instead of zero or infinity \cite{Majhi1}, we obtain, using (\ref{ansatz}) and (\ref{1.10}),
\begin{eqnarray}
{\textrm{Im}}~t = -{\textrm{Im}}\int\frac{dr}{\sqrt{f(r)g(r)}}.
\label{tunn3}
\end{eqnarray}
Therefore the outgoing probability is
\begin{eqnarray}
P_{{\textrm{out}}}=|\phi_{(\textrm{out})}|^2={\textrm{exp}}\Big[-\frac{4}{\hbar}\omega\Big(1+\sum_i\beta_i\frac{\hbar^i}{M^{2i}}\Big){\textrm{Im}}\int\frac{dr}{\sqrt{f(r)g(r)}}\Big].
\label{tunn4}
\end{eqnarray}
Then use of the principle of ``detailed balance'' \cite{Paddy,Bibhas} $P_{{\textrm{out}}}= {\textrm {exp}}\Big(-\omega\beta_{\textrm{(corr.)}}\Big)P_{\textrm{in}}={\textrm{exp}} \Big(-\omega\beta_{\textrm{(corr.)}}\Big)$ yields the corrected inverse Hawking temperature \cite{Majhi1} 
\begin{eqnarray}
\beta_{(\textrm{corr.})}=\beta_H\Big(1+\sum_i\beta_i\frac{\hbar^i}{M^{2i}}\Big),
\label{new1}
\end{eqnarray}
where $T_H=\beta_H^{-1}=\frac{\hbar}{8\pi M}$ is the standard semiclassical Hawking temperature.
The non leading terms are the corrections to temperature due to quantum effect. From the first law of thermodynamics $dS_{\textrm{bh}}=\beta_{(\textrm{corr.})}dM$ it is easy to find the corrected form of the Bekenstein-Hawking entropy which in this case is given by \cite{Majhi1}
\begin{eqnarray}
S_{\textrm{bh}}=\frac{4\pi M^2}{\hbar} + 8\pi\beta_1 \ln M-\frac{4\pi\hbar\beta_2}{M^2}+{\textrm{higher order terms in $\hbar$}}.
\label{new2}
\end{eqnarray} 
Expressing the above equation in terms of the usual semiclassical area $A=16\pi M^2$ yields,
\begin{eqnarray}
S_{\textrm{bh}}=\frac{A}{4\hbar} + 4\pi\beta_1 \ln A-\frac{64\pi^2\hbar\beta_2}{A}+{\textrm{higher order terms in $\hbar$}}
\label{area}
\end{eqnarray}
which is the corrected form of the Bekenstein-Hawking area law. The first term in (\ref{new2}) or (\ref{area}) is the usual semiclassical result while the second term is the logarithmic correction \cite{Majhi,Majhi1,Fursaev,Majumdar,Page} which in this case comes from $\hbar$ order correction to the action $S(r,t)$ and so on. In the next section we will discuss a method of fixing the coefficient $\beta_1$.

\section{Connection with trace anomaly}
      Naively one would expect $T_\mu^\mu$, the trace of the energy momentum tensor, to vanish for a zero rest mass field. However this is not the case since it is not possible to simultaneously preserve conformal and diffeomorphism symmetries at the quantum level. As the latter symmetry is usually retained there is, in general, a violation of the conformal invariance which is manifested by a non vanishing trace of the energy-momentum tensor. We now show that the coefficient $\beta_1$ appearing in (\ref{new2}) is related to this trace anomaly.

     We begin by studying the behaviour of the action, upto order $\hbar$, under an infinitesimal constant scale transformation, parametrised by $k$, of the metric coefficients,
\begin{eqnarray}
\bar{g}{_{\mu\nu}} = k g_{\mu\nu}\simeq(1+\delta k)g_{\mu\nu}.
\label{1.13}
\end{eqnarray}
Under this the metric coefficients of (\ref{1.1}) change as $\bar f=kf,\bar g=k^{-1}g$. Also, in order to preserve the scale invariance of the Klein-Gordon equation (\ref{1.2}) $\partial_\mu(\sqrt{-g}g^{\mu\nu}\partial_\nu)\phi=0$, $\phi$ should transform as $\bar\phi=k^{-1}\phi$. On the other hand, $\phi$ has the dimension of mass and since in our case the only mass parameter is the black hole mass $M$, the infinitesimal change of it is given by,
\begin{eqnarray}
\bar{M}=k^{-1}M\simeq(1-\delta k)M.
\label{1.16}
\end{eqnarray}
A similar result was also obtained by Hawking \cite{Hawking1} from other arguments.
 
      Now the form of the imaginary part of $S_0(r,t)$ for the outgoing particle, derived from (\ref{tunn11}) using (\ref{tunn3}) is given by,
\begin{eqnarray}
\textrm{Im}S_0^{(\textrm{out})}=-2\omega \textrm{Im}\int\frac{dr}{\sqrt{f(r)g(r)}}
\label{1.12}
\end{eqnarray}
where $\omega$ gets identified with the energy (i.e. mass $M$) of a stable black hole \cite{Majhi}. Therefore $\omega$ also transforms like (\ref{1.16}) under (\ref{1.13}).

       Considering only the $\hbar$ order term in (\ref{1.10}) and using (\ref{1.16}) we obtain, under the scale transformation,      
\begin{eqnarray}
\textrm{Im}\bar{S}_1^{(\textrm{out})}&=&\frac{\beta_1}{\bar{M}^2}\textrm{Im}\bar S_0^{(\textrm{out})}=-\frac{2\beta_1}{\bar{M}^2}{\bar{\omega}}{\textrm{Im}}\int\frac{dr}{\sqrt{{\bar{f}}{\bar{g}}}}
\nonumber
\\
&\simeq&-\frac{2\beta_1}{M^2(1-\delta k)}\omega{\textrm{Im}}\int\frac{dr}{\sqrt{fg}}\simeq\frac{\beta_1}{M^2}(1+\delta k)\textrm{Im}S_0^{(\textrm{out})}.
\label{1.14}
\end{eqnarray} 
Therefore,
\begin{eqnarray}
\delta{\textrm{Im}}S_1^{({\textrm{out}})}&=&{\textrm{Im}}{\bar{S_1}}^{({\textrm{out}})}-{\textrm{Im}}S_1^{({\textrm{out}})}
\nonumber
\\
&\simeq&\delta k\frac{\beta_1}{M^2}{\textrm{Im}}S_0^{({\textrm{out}})}
\label{delta}
\end{eqnarray} 
leading to,
\begin{eqnarray}
\frac{\delta \textrm{Im}S_1^{(\textrm{out})}}{\delta k}=\frac{\beta_1}{M^2}\textrm{Im}S_0^{(\textrm{out})}.
\label{1.15}
\end{eqnarray}
Now use of the definition of the energy-momentum tensor and (\ref{1.15}) yields,
\begin{eqnarray}
\textrm{Im}\int d^4x\sqrt{-g} T_\mu^\mu = \frac{2\delta \textrm{Im}S_1^{(\textrm{out})}}{\delta k}=\frac{2\beta_1}{M^2}\textrm{Im}S_0^{(\textrm{out})}.
\label{1.17}
\end{eqnarray}
Thus, in the presence of a trace anomaly, the action is not invariant under the scale transformation. This relation connects the coefficient $\beta_1$ with the trace anomaly. Since for the Schwarzschild black hole $f(r)=g(r)=1-\frac{2M}{r}$, from (\ref{1.12}) we obtain $\textrm{Im}S_0^{(\textrm{out})}=-4\pi\omega M$. Substituting this in (\ref{1.17}) $\beta_1$ can be expressed as
\begin{eqnarray}
\beta_1= -\frac{M}{8\pi \omega}\textrm{Im}\int d^4x\sqrt{-g}T_\mu^\mu.
\label{new3}
\end{eqnarray}
For a stable black hole, as mentioned below (\ref{1.12}), $\omega=M$ and the above equation simplifies to,
\begin{eqnarray}
\beta_1= -\frac{1}{8\pi}\textrm{Im}\int d^4x\sqrt{-g}T_\mu^\mu.
\label{new31}
\end{eqnarray}
Using this in (\ref{new2}) the leading correction to the semiclassical contribution is obtained,
\begin{eqnarray}
S_{\textrm{bh}}=\frac{4\pi M^2}{\hbar}-\Big(\textrm{Im}\int d^4x\sqrt{-g}T_\mu^\mu\Big)\ln M.
\label{new4}
\end{eqnarray}

        We now show that the above result is exactly equivalent to Hawking's \cite{Hawking1} original calculation by path integral approach based on zeta function regularization where he has modified the path integral by including the effect of fluctuations due to the presence of scalar field in the black hole spacetime. The path integral has been calculated under the saddle point approximation leading to the following expression for the logarithm of the partition function,
\begin{eqnarray}
\ln Z=-\frac{4\pi M^2}{\hbar}+\zeta(0)\ln M
\label{new5}
\end{eqnarray}
where the zero of the zeta function is given by \cite{Hawking1}
\begin{eqnarray}
\zeta(0)=-\textrm{Im}\int d^4x\sqrt{-g}T_\mu^\mu.
\label{new6}
\end{eqnarray}
The first term in (\ref{new5}) is the usual semiclassical contribution. The second term is a one loop effect coming from the fluctuation of the scalar field. From this expression the corrected entropy can be inferred,
\begin{eqnarray}
S_{\textrm{bh}}&=&\ln Z+\beta_H M
\nonumber
\\
&=&\frac{4\pi M^2}{\hbar}-\Big(\textrm{Im}\int d^4x\sqrt{-g}T_\mu^\mu\Big)\ln M
\label{new7}
\end{eqnarray}
which reproduces (\ref{new4}). As before, $\beta_H$ is the semiclassical result for the inverse Hawking temperature $\beta_H=\frac{8\pi M}{\hbar}$.

    To obtain an explicit value for $\beta_1$ (\ref{new31}) it is necessary to calculate the trace anomaly. For a scalar background this is given by \cite{Witt},
\begin{eqnarray}
T_\mu^\mu=\frac{1}{2880\pi^2}[R_{\mu\nu\alpha\lambda} R^{\mu\nu\alpha\lambda}-R_{\mu\nu}R^{\mu\nu}+\nabla_\mu\nabla^\mu R].
\label{1.19}
\end{eqnarray}
Inserting this in (\ref{new31}) yields,
\begin{eqnarray}
\beta_1&=& -\frac{1}{8\pi}\frac{1}{2880\pi^2}\textrm{Im}\int_{r=2m}^{\infty}\int_{\theta=0}^{\pi}\int_{\phi=0}^{2\pi}\int_{t=0}^{-8i\pi M}\frac{48 M^2}{r^6} r^2 {\textrm{sin}}\theta dr d\theta d\phi dt
\nonumber
\\
&=&\frac{1}{360\pi}.
\label{1.20}
\end{eqnarray}
As a check we recall that the general form for $\beta_1$ is given by \cite{Fursaev}
\begin{eqnarray}
\beta_1=-\frac{1}{360\pi}\Big(-N_0-\frac{7}{4}N_{\frac{1}{2}}+13 N_1+\frac{233}{4}N_{\frac{3}{2}}-212N_2\Big)
\label{conformal}
\end{eqnarray}
where `$N_s$' denotes the number of fields with spin `$s$'. The result (\ref{1.20}) is reproduced from (\ref{conformal}) by setting $N_0=1$ and $N_{\frac{1}{2}}=N_1=N_{\frac{3}{2}}=N_2=0$.

\section{Corrected Schwarzschild metric} 
      In this section, exploiting the corrected temperature and Hawking's periodicity arguments \cite{Hawking2}, the modification to the Schwarzschild metric will be calculated. The corrected inverse Hawking temperature is given by (\ref{new1}). Now if one considers this $\beta_{(\textrm{corr.})}$ as the new periodicity in the euclidean time coordinate $\tau$ , then following Hawking's arguments \cite{Hawking2} the euclidean form of the metric will be given by,
\begin{eqnarray}
ds^2_{(\textrm{corr.})}=x^2\Big[\frac{\displaystyle d\tau}{\displaystyle 4M\Big(1+\sum_i\beta_i\frac{\hbar^i}{M^{2i}}\Big)}\Big]^2+\Big(\frac{r^2}{r_h^2}\Big)^2dx^2+r^2d\Omega^2.
\label{1.22}
\end{eqnarray}      
This metric is regular at $x=0$, $r=r_h$ and $\tau$ is regarded as an angular variable with period $\beta_{(\textrm{corr.})}$. Here $r_h$ is the corrected event horizon for the black hole whose value will be derived later. Taking $f_{(\textrm{corr.})}(r)$ as the corrected metric coefficient for $d\tau^2$ we define a transformation of the form $x=4M\Big(1+\displaystyle \sum_i\beta_i\frac{\hbar^i}{M^{2i}}\Big)f_{(\textrm{corr.})}^{\frac{1}{2}}(r)$
under which the above metric simplifies to
\begin{eqnarray}
ds^2_{(\textrm{corr.})}= f_{(\textrm{corr.})}(r)d\tau^2+\frac{dr^2}{g_{(\textrm{corr.})}(r)}+r^2d\Omega^2
\label{1.24}
\end{eqnarray}
where the form of the $g_{(\textrm{corr.})}(r)$ is 
\begin{eqnarray}
g_{(\textrm{corr.})}(r)=\Big(\frac{r_h^2}{r^2}\Big)^2\frac{1}{(2M)^2}\Big(1+\sum_i\beta_i\frac{\hbar^i}{M^{2i}}\Big)^{-2}f_{(\textrm{corr.})}f'{_{(\textrm{corr.})}^{-2}}.
\label{1.25}
\end{eqnarray}
Now to get the exact form of these metric coefficients we will use the asymptotic limit: as $r\rightarrow\infty$, $f_{(\textrm{corr.})}(r)\rightarrow 1$ and $g_{(\textrm{corr.})}(r)\rightarrow 1$. With these boundary conditions $f'_{(\textrm{corr.})}(r)$ has the following general form,
\begin{eqnarray}
f'_{(\textrm{corr.})}(r)=\frac{r_h^2}{r^2}\frac{1}{2M}\Big(1+\sum_i\beta_i\frac{\hbar^i}{M^{2i}}\Big)^{-1}\Big(1+\sum_{n=1}^\infty C_nr^{-n}\Big)
\label{1.26}
\end{eqnarray}
where $C_n$~s are some constants. After integration 
\begin{eqnarray}
f_{(\textrm{corr.})}(r)=-\frac{r_h^2}{2M}\Big(1+\sum_i\beta_i\frac{\hbar^i}{M^{2i}}\Big)^{-1}\Big[\frac{1}{r}+\sum_{n=1}^\infty C_n\frac{r^{-(n+1)}}{n+1}\Big]+\textrm{constant}.
\label{1.27}
\end{eqnarray}
The asymptotic limit determines the integration constant as $1$. The constants $C_n$ will be determined by imposing the condition that when there is no quantum correction ($\hbar\rightarrow 0$) then this metric coefficient should reduces to its original form. This condition shows that each $C_n$ must be equal to zero. So the corrected metric coefficients are
\begin{eqnarray}
f_{(\textrm{corr.})}(r)=g_{(\textrm{corr.})}(r)=1-\frac{r_h^2}{2Mr}\Big(1+\sum_i\beta_i\frac{\hbar^i}{M^{2i}}\Big)^{-1}.
\label{1.28}
\end{eqnarray}
For a static black hole the event horizon is given by $g_{tt}(r_h)=g^{rr}(r_h)=0$. Therefore in this case the event horizon is given by
\begin{eqnarray}
r_h=2M\Big(1+\sum_i\beta_i\frac{\hbar^i}{M^{2i}}\Big).
\label{1.29}
\end{eqnarray}
Finally, inserting (\ref{1.29}) in (\ref{1.28}) and changing $\tau\rightarrow it$ lead to the corrected form of the metric (\ref{1.24}) as, 
\begin{eqnarray}
ds^2_{(\textrm{corr.})}=-\Big[1-\frac{2M}{r}\Big(1+\sum_i\beta_i\frac{\hbar^i}{M^{2i}}\Big)\Big]dt^2+\frac{\displaystyle dr^2}{\displaystyle\Big[1-\frac{2M}{r}\Big(1+\sum_i\beta_i\frac{\hbar^i}{M^{2i}}\Big)\Big]}+r^2d\Omega^2. 
\label{1.30}
\end{eqnarray}
This modified metric includes all quantum corrections. Expectedly for $\hbar\rightarrow 0$ it reduces to the standard Schwarzschild metric.

      A discussion on the comparison of the above results with the earlier works \cite{York,Lousto} is feasible. If we confine ourself to ${\cal{O}}(\hbar)$ only, with $\beta_1$ defined by (\ref{new31}), then equations(\ref{1.29}) and (\ref{1.30})reduce to
\begin{eqnarray}
r_h=2M\Big(1+\beta_1\frac{\hbar}{M^{2}}\Big)
\label{1.31}
\end{eqnarray}
and
\begin{eqnarray}
ds^2_{(\textrm{corr.})}=-\Big[1-\frac{2M}{r}\Big(1+\beta_1\frac{\hbar}{M^{2}}\Big)\Big]dt^2+\frac{\displaystyle dr^2}{\displaystyle\Big[1-\frac{2M}{r}\Big(1+\beta_1\frac{\hbar}{M^{2}}\Big)\Big]}+r^2d\Omega^2. 
\label{1.32}
\end{eqnarray}
These have a close resemblance with the results obtained before \cite{York,Lousto} by solving the semiclassical Einstein equations containing the one loop renormalized energy-momentum tensor. This tensor acts as a source of curvature (back reaction effect) modifying the metric and the horizon radius by terms proportional to the trace anomaly, analogous to (\ref{1.31}, \ref{1.32}).
\\

\section{Summary and discussions}           
     Let us summarise our findings. We gave a new approach to get the quantum Hamilton-Jacobi equation from the definition of the quantum canonical momentum. Using this equation and adopting the tunneling method \cite{Majhi1}, the corrected form of the inverse Hawking temperature was obtained. Exploiting the first law of thermodynamics, the logarithmic and inverse powers of area corrections to the Bekenstein-Hawking area law were reproduced. Arguments based on a constant scale transformation to the metric revealed that the coefficient of the logarithmic correction was proportional to the trace anomaly. The explicit value of this coefficient was computed in the case of a scalar background. An exact equivalence of our ${\cal{O}}(\hbar)$ result with Hawking's \cite{Hawking1} original result obtained by computing the one loop effective action employing zeta function regularization was discussed. Finally, the modification to the original Schwarzschild metric was derived on the basis of the corrected temperature and periodicity arguments \cite{Hawking2}. Upto ${\cal{O}}(\hbar)$ correction, this modification was similar to the existing results \cite{York,Lousto} obtained by including one loop back reaction effect.

    We observe, therefore, that the equivalence between Hawking's original calculations \cite{Hawking1} and the tunneling formalism in the Hamilton-Jacobi approach is valid even beyond the semiclassical approximation. In the original scheme the one loop effects were computed from a zeta function regularization of the path integral obtained from fluctuations of a scalar field. Here, on the other hand, these effects were obtained from a modification, beyond the usual WKB approximation, in the one particle action corresponding to a massless scalar particle.

   Let us next discuss the precise connection between Hawking's analysis \cite{Hawking1} and the tunneling formalism. We feel that the tunneling mechanism adapted by us is closest in spirit to the original computations \cite{Hawking1}, thereby explaining the equivalence of the results. The point is that formal considerations \cite{Morette} imply, upto ${\cal{O}}(\hbar)$, an equivalence of the path integral with the WKB ansatz $\phi={\textrm{exp}}[-\frac{i}{\hbar}S]$ adopted here. What we established was the precise equivalence of the ${\cal{O}}(\hbar)$ one particle action with the path integral obtained by the zeta function regularization approach. As far as the evaluation of the one particle action was concerned, no specific regularization was necessary. For path integrals, however, meaningful expressions can only be abstracted by using an appropriate regularization. In this sense the zeta function regularization of path integrals in curved background was singled out. This view is compatible with \cite{Hawking1} where it was shown that other regularizations- like dimensional regularization- led to ambiguous results.

    We would like to mention that non thermal corrections to the Hawking effect were earlier discussed in \cite{Parikh} following a different approach. Usually the tunneling probability takes the form $\Gamma\sim e^{-\beta_H\omega}$ from which the Hawking temperature is identified as $\beta_H^{-1}$. Keeping energy conservation, $\Gamma$ was shown to behave as $e^{-\beta_H\omega(1-\frac{\omega}{2M})}$ thereby deviating the spectrum from thermality. In our case the structure of $\Gamma$ is retained as $e^{-\beta_{(\textrm{corr.})}\omega}$, where $\beta_{(\textrm{corr.})}$ is given by (\ref{new1}).

     As a final remark we feel that, although the results are presented here for a scalar background, the methods are applicable for other cases.


\end{document}